\documentclass[aps,pra,twocolumn,showpacs,superscriptaddress,groupedaddress]{revtex4-2}  
\pdfoutput=1
\usepackage{graphicx}  
\usepackage{dcolumn}   
\usepackage{bm}        
\usepackage{amssymb}   

\usepackage{dsfont}
\usepackage{amsmath}
\usepackage{bbm}
\usepackage{here}
\usepackage{hyperref}
\usepackage{qcircuit}
\usepackage{setspace}

\usepackage{lipsum}

\begin{document}

\newcommand{\ket}[1]{\left| #1 \right\rangle}
\newcommand{\bra}[1]{\left\langle #1 \right|}
\newcommand{\cA}{\mathcal{P}}
\newcommand{\cH}{\mathcal{H}}
\newcommand{\cL}{\mathcal{L}}
\newcommand{\cK}{\mathcal{K}}
\newcommand{\cM}{\mathcal{M}}
\newcommand{\cP}{\mathcal{P}}
\newcommand{\cR}{\mathcal{R}}
\newcommand{\cS}{\mathcal{S}}
\newcommand{\cU}{\mathcal{U}}

\newcommand{\bracket}[2]{\langle {#1}|{#2} \rangle}
\newcommand{\ketbra}[2]{\ket{#1}\negmedspace\bra{#2}}
\newcommand{\id}{\mathbf{1}}


\title{
Robust Micro-Macro Entangled States}
                            
\author{Maryam Sadat Mirkamali}
 \email{msmirkam@uwaterloo.ca}
\affiliation{Department of Physics, Sharif University of technology, PO Box 11155-9161, Tehran, Iran}
\author{David G. Cory}%
\affiliation{Institute for Quantum Computing, University of Waterloo, Waterloo, Ontario N2L 3G1, Canada}
\affiliation{Department of Chemistry, University of Waterloo, Waterloo, Ontario  N2L 3G1, Canada}

\date{\today}

\begin{abstract}

Bipartite entangled states between a qubit and macroscopically distinct states of a mesoscopic system, known as micro-macro entangled states, are emerging resources for quantum information processing. 
One main challenge in generating such states in the lab is their fragility to environmental noise.
We analyse this fragility in details for single particle noise by identifying what factors play a role in the robustness and quantifying their effect.
 There is a trade off between the macroscopicity of a micro-macro entangled state and the robustness of its bipartite entanglement to environmental noise. 
We identify symmetric micro-macro entangled states as the most robust states to single particle noise.
We show that the robustness of bipartite entanglement of such states to single particle noise 
decreases as the second order of macroscopicity, which identifies a regime where the bipartite entangled state is both robust and macroscopic. Our result is a step towards retaining quantum characteristics in large scales and experimental realization of micro-macro entangled spin states and their use for connecting separated qubits. Moreover, it advances our understanding of quantum to classical transition. 
 
\end{abstract}

\maketitle


\section{Introduction}
\label{sec:Intro}

Micro-macro entangled states and more generally superposition of macroscopically distinct states, known as macroscopic superposition states, are of both fundamental and practical interests. 
They have been used to formulate fundamental
questions about quantum mechanics such as: 
How quantum mechanics applies to large scales? What causes quantum to classical transition? and
What is a valid interpretation of quantum superposition and projective quantum measurement based on Schrodinger's thought cat experiment? \cite{Schrodinger35,Zurek03,Leggett02}
With advances in quantum technologies, micro-macro entangled states between a qubit and a mesoscopic spin system consisting of two-level particles are
identified as useful resources for quantum information processing applications such as connecting separated qubits  
\cite{Mirkamali18}.
There are proposals for creating micro-macro entangled spin states experimentally \cite{Mirkamali20}.
When it comes to experimental realization robustness to imperfections is an
essential aspect to be considered.

Here, we analyse the robustness of micro-macro entangled states between a qubit and a mesoscopic system consisting of two-level particles (e.g. spin-half particles) to microscopic events including single particle measurement and single particle loss \footnote{Particle loss for spin system is equivalent to complete depolarization and can be a model of $T_1$ relaxation.}.
Micro-macro entanglement is identified by two independent features: (1) bipartite entanglement and (2) macroscopic distinctness or macroscopicty (to be defined precisely below). 
Among these two properties macroscopic distinctness is, by definition, robust to microscopic noise.
We analyse the robustness of bipartite entanglement to microscopic events. In particular, we answer the following questions: (1) How bipartite entanglement of a micro-macro entangled state varies as a function of macroscopicity upon single particle loss and single particle measurement? 
(2) Apart from macroscopicity, what other 
properties of a micro-macro entangled state play role in its robustness to microscopic noise? 
(3) What is the upper bound on robustness as a function of macroscopicity? and (4) among all micro-macro entangled states which ones are the most robust to microscopic noise?


Bipartite entanglement in a micro-macro entangled state is in analogy with coherence in a macroscopic superposition state.
Coherence of the macroscopic superposition states is known to be fragile to noise \cite{Frowis18}; up to the point that some proposed measures of macroscopicity of these states are based on how fragile they are to a particular form of noise \cite{Dur02,Lee2011,Kwon17,Sekatski14}.
In this work, we quantify macroscopicity by distinguishability with a physical collective observable which is the measure proposed by Leggett \cite{Leggett02}. In this sense, our work relates the measure of macroscopicity based on Leggett's proposal to a measure based on robustness to microscopic noise.


We show that the robustness of the entanglement of a micro-macro entangled state between a qubit and a mesoscopic spin system reduces with its macroscopicity; consistent with the fragility of macroscopic superposition states to noise \cite{Zurek03b}. 
We prove that for symmetric micro-macro entangled states, the initial drop in the entanglement
after single particle loss 
is second
order in the distinctness. 
As a result, we identify a regime where micro-macro entangled states are
both robust and macroscopic.
Furthermore, we prove that micro-macro entangled states that are symmetric with respect to their preferred axis are the most robust states to single particle noise. Thus apart from  distinctness, symmetry of the micro-macro entangled states is an essential factor in robustness.  


This result is of great practical importance. It has been shown that the usefulness of micro-macro entangled states in quantum information processing, for connecting separated qubits, increases with the distinctness \cite{Mirkamali20}. Our analysis confirms the trade off between usefulness and robustness of micro-macro entangled states. It also guides us to find micro-macro entangled states that are both robust and useful.
In addition to practical relevance, it improves our understanding of quantum behaviour at large scales and quantum to classical transition in many-body quantum systems.

\section{Statement of the Problem}
\label{sec:microscopic}

Micro-macro entangled states are identified with two features: 1. bipartite entanglement between a microscopic system such as a qubit and a many-body system like a mesoscopic spin system and 2. macroscopic distinctness between the states of the many-body system that are correlated with the state of the microscopic system. These two characteristics make micro-macro entangled states involving mesoscopic \textit{spin} systems (MSSs) a resource for connecting separated qubits (while having access to limited control) \cite{Mirkamali20}. Here 'spin' refers to any two-level quantum system, such as spin-half particles, Rydberg atoms, trapped ions, etc. In the rest of this paper, without loss of generality, we consider spin half particles as an example. 

Micro-macro entangled states are a subset of macroscopic superposition states, which are defined as superposition of two macroscopically distinct states, usually represented with $\ket{A}$ (Alive) and $\ket{D}$ (Dead), $\frac{1}{\sqrt{2}}\left(\ket{A}+\ket{D}\right)$ \cite{Frowis18}. 

A micro-macro entangled state has a general form of,
\begin{equation}
\label{eq:mic-mac}
    \ket{\phi}_{q-\text{MSS}}=\frac{1}{\sqrt{2}}\left(\ket{0}_{q}\ket{\psi_0}_{\text{MSS}}+\ket{1}_{q}\ket{\psi_1}_{\text{MSS}}\right)
\end{equation}
Bipartite entanglement requires the states $\ket{\psi_0}_{\text{MSS}}$ and $\ket{\psi_1}_{\text{MSS}}$ to be orthogonal, $\bracket{\psi_1}{\psi_0}_{\text{MSS}}=0$. 
There 
is no unique definition for macroscopicity or macroscopic distinctness between $\ket{\psi_0}_{\text{MSS}}$ and $\ket{\psi_1}_{\text{MSS}}$ in a micro-macro entangled state or $\ket{A}$ and $\ket{D}$ in a macroscopic superposition state \cite{Leggett02,Dur02,Shimizu2002,Bjork04,Cavalcanti06,Cavalcanti08,Korsbakken2007,Marquardt08,Lee2011,Frowis12,Nimmrichter13,Sekatski14,Sekatski18,Laghaout15,Yadin15,Kwon17}. 
Here, we choose a measure of macroscopic distinctness that follows usefulness in connecting separated qubits \cite{Mirkamali20,Leggett02} and is based on distinguishability with a coarse-grained low-resolution collective 
measurement. The same measure has been used to experimentally quantify macroscopicity of micro-macro entangled states of photons \cite{Lvovsky13,Bruno13}.

\begin{figure}[t!h]
\centering
         \includegraphics[scale=0.8]{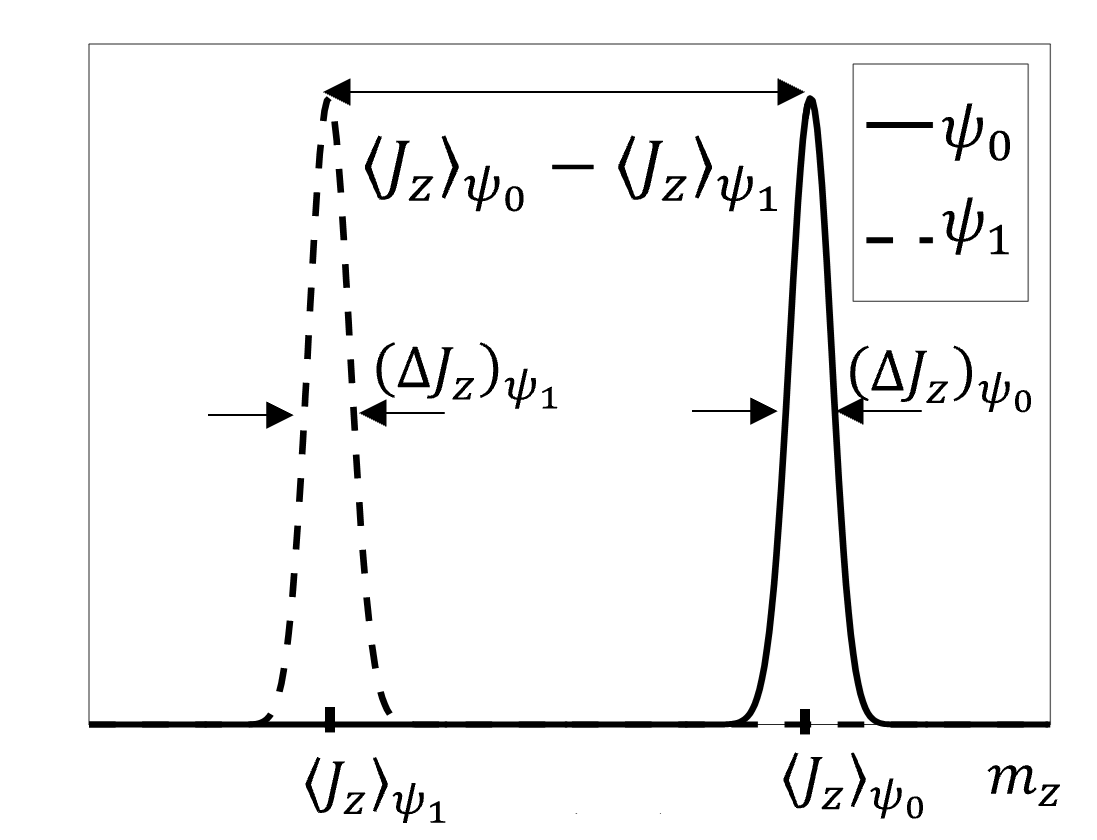}
                 \caption{\label{fig:mic-mac_macroscopic} Schematic representation of the magnetization spectrum of a typical micro-macro entangled spin state. Macroscopic distinctness requires the spectra of $\ket{\psi_0}_{\text{MSS}}$ and $\ket{\psi_1}_{\text{MSS}}$ to be separated such that the two states are distinguishable with a coarse-grained low-resolution collective measurement. Note that the magnetization spectrum of any state is discrete. This continuous plot shows the envelope of the spectrum. }    
\end{figure}

To describe what macroscopic distinctness means, we first define the distinctness. 
 The distinctness of the state \ref{eq:mic-mac}, or the the "extensive difference" as called by Leggett \cite{Leggett02}, is defined as  the difference in the expectation values for the two states  $\ket{\psi_0}_{\text{MSS}}$ and $\ket{\psi_1}_{\text{MSS}}$ (or $\ket{A}$ and $\ket{D}$ in a macroscopic superposition state) maximized over all extensive physical observables.
For a mesoscopic system consisting of spin-half particles, this extensive observable can be collective magnetization along some preferred axis, $J_z$. This preferred axis is specified by the micro-macro entangled state, itself. It is a direction along which the two component states of the mesoscopic system,$\ket{\psi_0}_{\text{MSS}}$ and $\ket{\psi_1}_{\text{MSS}}$, show extensive difference. 
We will call this axis the quantization axis (QA) or the measurement axis (MA) and we will denote it by $z$. 
The distinctness of the state \ref{eq:mic-mac} is then, mathematically, defined as,

 \begin{equation}
 \label{eq:macroscopicDis}
 \Lambda(\phi) =|\left\langle J_z \right\rangle_{\psi_0}- \left\langle J_z \right\rangle_{\psi_1}|
 \end{equation}
 A bipartite entangled state 
 $\ket{\phi}$ is macroscopically distinct or macroscopic, in short, if  $\Lambda(\phi)$ is large compared to the quanta of collective magnetization, $\hbar$, the standard deviation of the collective magnetization for both of the states, $\left(\Delta J_z\right)_{\psi_0}$ and $\left(\Delta J_z\right)_{\psi_1}$, and the resolution of the measurement apparatus. To have distinguishable states, it is also required that there is no overlap of the magnetization spectra, 
$P(m_z)_{\psi_0}P(m_z)_{\psi_1}=0$ ($P(m_z)_{\psi_0}=\bra{\psi_0}\Pi(m_z)\ket{\psi_0}$ where $\Pi(m_z)$ is the projection operator to subspace whose magnetization is $m_z$. $P(m_z)_{\psi_1}$ is defined similarly) and the magnetization spectra are extended over non-overlapping distinct ranges, as depicted in figure \ref{fig:mic-mac_macroscopic}.

Among the two characteristics of micro-macro entangled states macroscopic distinctness is 
expected (and, here, will be confirmed,) to be robust to microscopic events, while the entanglement is fragile. Generally the more macroscopic the distinctness is, the more fragile the entanglement is. 
The main question that this paper answers is formulated as: How robustness of entanglement in a micro-macro entangled state to microscopic events (including single particle measurement and single particle loss) varies with its 
distinctness?
We use entanglement entropy \cite{Horodecki09,HORODECKI94} and negativity \cite{Vidal02} for quantifying entanglement after single particle measurement and single particle loss, respectively, and $\Lambda(\phi)$ in equation \ref{eq:macroscopicDis} as the measure of distinctness.
The next questions covered here are: How other factors such as symmetry of the state and standard deviation of its spectra play a role in the fragility of entanglement? What is the upper bound on the robustness of bipartite entanglement as a function of distinctness? What micro-macro states are the most robust to microscopic events?
Answering these questions is crucial from both fundamental and practical 
points of view. It improves our understanding of the quantum to classical transition in many-body quantum systems. It also helps identifying requirements for experimental realization of micro-macro entangled spin states. 

The rest of this paper is organized as following.
First, we discuss micro-macro entangled states where the corresponding states of the mesoscopic spin system, $\ket{\psi_0}_{\text{MSS}}$ and $\ket{\psi_1}_{\text{MSS}}$ are idealized Dicke states (section \ref{sec:Dicke}). Dicke states have two properties that make their study insightful: (1) Their magnetization spectra along the QA has a single value i.e. their SD is zero which helps analysing the effect of distinctness (difference in the mean). (2) They are symmetric with respect to swapping their constituent spins, which makes the effect of the noise independent of the spin index. 
Next, we relax the limiting properties of Dicke states and generalize the analysis to all micro-macro entangled states whose mesoscopic spin system states'  $\ket{\psi_0}_{\text{MSS}}$ and $\ket{\psi_1}_{\text{MSS}}$ are symmetric with respect to the quantization axis or the measurement axis; meaning that measuring any single spin in the MSS along the QA results in similar statistics  
(section \ref{sec:symmetric}). The states, of course, need to satisfy the macroscopic distinctness condition but their spectra can be extended over a range of magnetization. 
We show that for Dicke states and more generally for these generalized 
QA-symmetric states the decrease in the entanglement after single particle measurement/loss is second order in the distinctness. Then, we will show that the 
QA-symmetric states are the most robust micro-macro entangled states to single particle measurement or single particle loss. 
\textit{ In other worlds, the most robust states are the ones that all spins in the mesoscopic system contribute
equally to the entanglement and to the macroscopic distinctness.}
We will demonstrate an upper bound 
for entanglement of micro-macro entangled states subjected to particle measurement/loss (\ref{sec:Bounds}). 


\section{Dicke States}
\label{sec:Dicke}



As a first step, we study robustness of micro-macro entangled states whose mesoscopic systems are in the idealized symmetric Dicke states. 
We should emphasize that we are not suggesting creation of the Dicke states in the lab. 
The motivation for this study is that the analysis of Dicke states is insightful and easy to follow. It helps seeing the effects of different factors, including mean of the spectra, the SD of the spectra, and distribution of the information among the spins, separately and it leads to analysing experimentally relevant states, in the next section.
 
Consider a micro-macro entangled state
with the mesoscopic side in the Dicke states
\begin{equation}
\label{eq:mic-mac-Dicke}
    \ket{\phi^D}_{q-\text{MSS}}=\frac{1}{\sqrt{2}}\left(\ket{0}_{q}\ket{D_{N,k_0}}+\ket{1}_{q}\ket{D_{N,k_1}}\right)
\end{equation}
where $\ket{D_{N,k_j}}$ is a symmetric Dicke state with $N$ spin-half particles with $k_j$ of them in state $\ket{\uparrow}$,
\begin{equation}
    \ket{D_{N,k_j}}=\frac{1}{\sqrt{\binom{N}{k_j}}}\sum_i^{\binom{N}{k_j}} P_i\left(\ket{\uparrow}^{\otimes k_j}\ket{\downarrow}^{\otimes N-k_j} \right)
\end{equation}
for $j=0,1$. $P_i$ is the permutation operator and the summation is over all permutations with $k_j$ spins $\ket{\uparrow}$ and $N-k_j$ spins $\ket{\downarrow}$. 
The state $\ket{\phi^D}_{q-\text{MSS}}$ is a maximally entangled state between the qubit and the MSS for all 
$k_0\neq k_1$.

For any Dicke state, $\ket{D_{N,k}}$, the expectation value of the collective magnetization along the quantization axis is $\left\langle J_z \right\rangle_{D_{N,k}}=\hbar\left(k-\frac{N}{2} \right)$ and the standard deviation is zero. 
  Thus, setting $\hbar$ to $1$, the distinctness of the micro-macro entangled state $\ket{\phi^D}_{q-\text{MSS}}$ is $\Lambda(\phi^D)=|\Delta k|=|k_0-k_1|$.

Upon single particle measurement along $z$, the joint state of the qubit and the MSS projects into one of the following states, $\ket{\phi^{D\uparrow}_{q,\text{MSS}}}$ or $\ket{\phi^{D\downarrow}_{q,\text{MSS}}}$ with their corresponding probability, $P_{\uparrow}$ and $P_{\downarrow}$,

\begin{widetext}
\begin{equation}
\label{eq:updatedDicke}
\begin{cases} 
\ket{\phi^{D\uparrow}_{q,\text{MSS}}}=\sqrt{\frac{k_0}{k_0+k_1}}\ket{0}_q\otimes\ket{\uparrow}\ket{D_{N-1,k_0-1}}+\sqrt{\frac{k_1}{k_0+k_1}}\ket{1}_q\otimes\ket{\uparrow}\ket{D_{N-1,k_1-1}}, \hspace{1.3cm} P_{\uparrow}=\dfrac{k_0+k_1}{2N}\\
\ket{\phi^{D\downarrow}_{q,\text{MSS}}}=\sqrt{\frac{N-k_0}{2N-(k_0+k_1)}}\ket{0}_q\otimes\ket{\downarrow}\ket{D_{N-1,k_0}}+\sqrt{\frac{N-k_1}{2N-(k_0+k_1)}}\ket{1}_q\otimes\ket{\downarrow}\ket{D_{N-1,k_1}}, \hspace{0.3cm} P_{\downarrow}=\dfrac{2N-(k_0+k_1)}{2N}
\end{cases}
\end{equation}
\end{widetext}
For simpler representation and without loss of generality, we considered measurement of the first spin in the MSS. 

\begin{figure}[t!h]
\centering
         \includegraphics[scale=0.42]{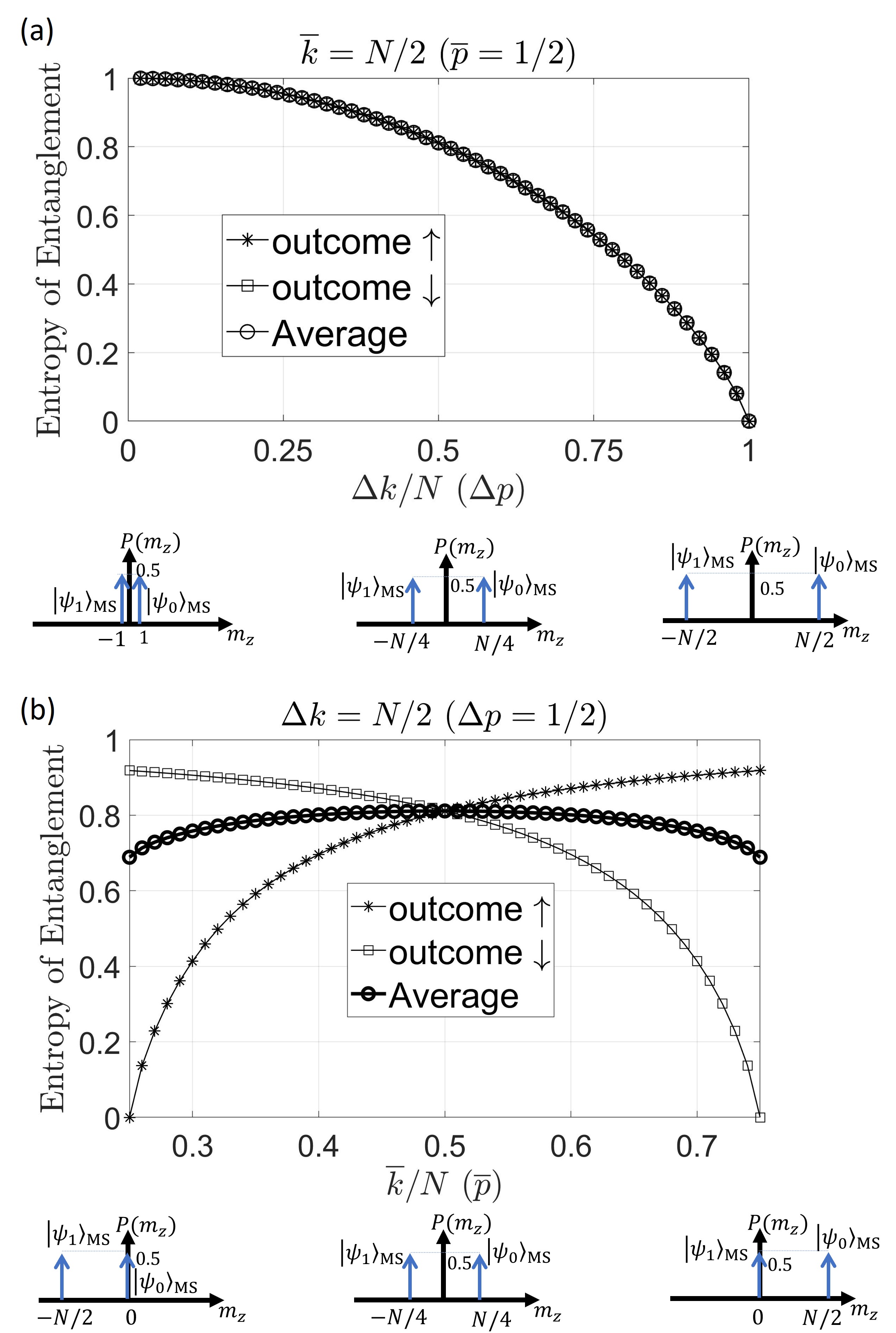}

                 \caption{\label{fig:Dicke_ent} 
Entropy of entanglement of a bipartite entangled state between a qubit and Dicke states (QA-symmetric states) of a mesoscopic spin system after single particle measurement. Plots (a) and (b) show entanglement as a function of distinctness and mean of the spectra, respectively, while the other parameter is fixed. 
                 The distinctness plays the main role in the robustness while the mean of the spectra has little effect.}    
\end{figure}

It is immediately seen that the distinctness is preserved for both of the states. 
How about the entanglement?
The entropy of entanglement of the updated states, quantified with Von Neumann entropy is,
\small
\begin{eqnarray}
\label{eq:EvDicke}
E_V\left(\phi^{D\uparrow}\right)&=&-\frac{k_0}{k_0+k_1} \log_2(\frac{k_0}{k_0+k_1})\\
&&-\frac{k_1}{k_0+k_1} \log_2(\frac{k_1}{k_0+k_1} ) \nonumber \\
E_V\left(\phi^{D\downarrow}\right)&=&-\frac{N-k_0}{2N-(k_0+k_1)} \log_2(\frac{N-k_0}{2N-(k_0+k_1)}) \nonumber \\
&&-\frac{N-k_1}{2N-(k_0+k_1)} \log_2(\frac{N-k_1}{2N-(k_0+k_1)}) \nonumber
\end{eqnarray}
\normalsize
It is insightful to plot the entanglement of the updated states once as a function of the distinctness, $\frac{\Lambda(\phi^D)}{N}=\frac{|k_0-k_1|}{N}$, and once as a function of the center of the spectra, $\frac{k_0+k_1}{2N}$, normalized to the size of the MSS, while the other quantity is fixed.
Figure \ref{fig:Dicke_ent} presents $E_V\left(\phi^{D\uparrow}\right)$ and $E_V\left(\phi^{D\downarrow}\right)$ and their average according to their corresponding probabilities, as a function of $\Delta k/N$ when $\frac{\Bar{k}}{N}=\frac{k_0+k_1}{2N}=1/2$ and as a function of  
$\frac{\Bar{k}}{N}$ when $\frac{\Delta k}{2N}=1/2$.

These plots show three main features.
First, they show that robustness of the entanglement to the single particle measurement is mostly affected by the distinctness and the centre of the spectra has little effect on the average entanglement.
Second, there is a trade off between macroscopic distinctness in a micro-macro entangled state and robustness of its bipartite entanglement to particle loss. This result is aligned with the fragility of the coherence in a macroscopic superposition state to noise 
\cite{Frowis18}.
Third, the subplot (a) shows that as the distinctness increases, the decrease in entropy of entanglement is first slow and then fast to reach the complete entanglement loss for $\Delta k=N$ which corresponds to the GHZ state with maximum distinctness.

The slow initial decrease is second order in $\Delta k$ for both $E_V\left(\phi^{D\uparrow}\right)$ and $E_V\left(\phi^{D\downarrow}\right)$. If $\Bar{k}=\frac{N}{2}$, the entropy of entanglement is,
\small
\begin{equation}
    \label{eq:second-order-Ev}
    E_V\left(\phi^{D\uparrow/\downarrow}\right)=
    1-\frac{1}{2\ln{2}}
    \left(\frac{\Delta k}{N}\right)^2+
    O\left(\left(\dfrac{\Delta k}{N}\right)^4\right)
\end{equation}
\normalsize
More generally when $\Bar{k}=\frac{k_0+k_1}{2}\neq \frac{N}{2}$, the next term in the expansion of the average Entropy of Entanglement is of the order of  $\left(\dfrac{\Delta k}{N}\right)^2 \left( \dfrac{\Bar{k}}{N}-\dfrac{1}{2}\right)^2$,
\small
\begin{eqnarray}
 \Bar{E}_V &=& P_{\uparrow}E_V\left(\phi^{D\uparrow}\right)+P_{\downarrow}E_V\left(\phi^{D\downarrow}\right) \\
  &=& 1-\frac{1}{2\ln{2}}\left(\frac{\Delta k}{N}\right)^2\left(1+ 4\left(\dfrac{\Bar{k}}{N}-\dfrac{1}{2}\right)^2 + \frac{1}{6} \left(\frac{\Delta k}{N}\right)^2+... \right) \nonumber
\end{eqnarray}
\normalsize


Dicke states are too specific and non-physical. In the next section, we show that these results; namely, dependency of robustness on distinctness and the second order initial drop, are quite general.
Any micro-macro entangled state for which all spins in the mesoscopic states, $\ket{\psi_0}$ and $\ket{\psi_1}$, have equal probability of being in state $\ket{\uparrow}$ or $\ket{\downarrow}$ behaves similarly upon single particle measurement. Moreover, these results are not limited to single particle measurement. The entanglement of such symmetric states after single particle loss, quantified by negativity, also depends mainly on distinctness with a second order initial drop.




\section{Collective QA-Symmetric States}
\label{sec:symmetric}
Consider 
a micro-macro entangled state 
\begin{equation}
\label{eq:mic-mac-Sym}
    \ket{\phi^S}_{q-\text{MSS}}=\frac{1}{\sqrt{2}}\left(\ket{0}_{q}\ket{S_{0}}+\ket{1}_{q}\ket{S_{1}}\right)
\end{equation}
with $\bracket{S_0}{S_1}=0$ and $\left<J_z\right>_{S_0}=M_0$ and $\left<J_z\right>_{S_1}=M_1$; 
where each of the states $\ket{S_{0}}$ and $\ket{S_{1}}$ are symmetric with respect to the preferred quantization axis (z-axis) meaning that measuring any particle in the MSS along the quantization axis yields the same mean value; 
Mathematically for any spin in the MSS, with index $i$, $\left<\sigma^i_z\right>_{S_0}=\frac{M_0}{N}$ and $\left<\sigma^i_z\right>_{S_1}=\frac{M_1}{N}$, where $N$ is the size of the mesoscopic system. 
We call this state a collective 
QA-symmetric micro-macro entangled state. It is collective since all spins in the MSS equally contribute in the distinctness (or are equally informed about the state of the qubit). 
It is 
QA-symmetric because for both $\ket{S_{0}}$ and $\ket{S_{1}}$ measuring any particle in the MSS along the quantization/measurement axis gives the same statistics. 


QA-symmetric states are an important subset of micro-macro entangled states which include experimentally relevant ones.
The micro-macro entangled spin states proposed to be generated with a mesoscopic spin system using experimentally available control tools \cite{Mirkamali20} are confirmed to be QA-symmetric, upto dynamical fluctuations. Another example of a QA-symmetric state is a micro-macro entangled state where   $\ket{S_{0}}$ and $\ket{S_{1}}$ are separable tensor products of the same superposition states for all spins:
$\ket{S_{0}}=\left( \cos\theta_0 \ket{\uparrow}+ \sin \theta_0 \ket{\downarrow} \right)^{\otimes N}$ and $\ket{S_{1}}=\left( \cos\theta_1 \ket{\uparrow}+ \sin \theta_1 \ket{\downarrow} \right)^{\otimes N}$. Note that the overlap of these two states is  $\bracket{S_0}{S_1}= \left(\cos(\theta_0-\theta_1)\right)^N$, and can be made arbitrarily small by enlarging the mesoscopic system or choosing separated enough $\theta_0$ and $\theta_1$. The magnetization spectra of $\ket{S_{0}}$ and $\ket{S_{1}}$ are Binomial distributions with mean and the SD of $M_j=\left\langle J_z\right\rangle_{S_j}=N/2(\cos \theta_j-\sin \theta_j)$ and $\left(\Delta J_z \right)_{S_j}= \sqrt{N \cos\theta_j \sin\theta_j}$ for $j=0,1$.

We show that for a collective 
QA-symmetric micro-macro entangled state, the amount of entanglement after single particle measurement or single particle loss depends only on the mean of the magnetization spectra normalized to the size of the system: $M_0/N$ and $M_1/N$. Thus their robustness does not depend on the detailed form of their state or higher moments of their magnetization spectra such as their SD.
  
Following a similar approach to the analysis of the Dicke states, after single particle measurement the updated states of a collective 
QA-symmetric micro-macro entangled state are
\begin{widetext}
\begin{equation}
\label{eq:updatedSym}
\begin{cases} 
\ket{\phi^{S\uparrow}_{q,\text{MSS}}}=\sqrt{\frac{p_0^{\uparrow}}{p_0^{\uparrow}+p_1^{\uparrow}}}\ket{0}_q\otimes\ket{\uparrow}\ket{S_{0,\uparrow}^{N-1}}+\sqrt{\frac{p_1^{\uparrow}}{p_0^{\uparrow}+p_1^{\uparrow}}}\ket{1}_q\otimes\ket{\uparrow}\ket{S_{1,\uparrow}^{N-1}}, \hspace{1.4cm} P_{\uparrow}=\dfrac{p_0^{\uparrow}+p_1^{\uparrow}}{2}\\
\ket{\phi^{S\downarrow}_{q,\text{MSS}}}=\sqrt{\frac{1-p_0^{\uparrow}}{2-(p_0^{\uparrow}+p_1^{\uparrow})}}\ket{0}_q\otimes\ket{\downarrow}\ket{S_{0,\downarrow}^{N-1}}+\sqrt{\frac{1-p_1^{\uparrow}}{2-(p_0^{\uparrow}+p_1^{\uparrow})}}\ket{1}_q\otimes\ket{\downarrow}\ket{S_{1,\downarrow}^{N-1}}, \hspace{0.3cm} P_{\downarrow}=\dfrac{2-(p_0^{\uparrow}+p_1^{\uparrow})}{2}
\end{cases}
\end{equation}
\end{widetext}
Assuming that the states in each set of  $\{\ket{S_{0\uparrow}^{N-1}},\ket{S_{1\uparrow}^{N-1}}\}$ and $\{\ket{S_{0\downarrow}^{N-1}},\ket{S_{1\downarrow}^{N-1}}\}$, are mutually orthogonal which follows from separation of the spectra in the original micro-macro entangled state, the entropies of entanglement of the above states are,
\small
\begin{eqnarray}
\label{eq:EvS}
E_V\left(\phi^{S\uparrow}\right)&=&-\frac{p_0^{\uparrow}}{p_0^{\uparrow}+p_1^{\uparrow}} \log_2(\frac{p_0^{\uparrow}}{p_0^{\uparrow}+p_1^{\uparrow}})\\
&&-\frac{p_1^{\uparrow}}{p_0^{\uparrow}+p_1^{\uparrow}} \log_2(\frac{p_1^{\uparrow}}{p_0^{\uparrow}+p_1^{\uparrow}}) \nonumber \\
E_V\left(\phi^{S\downarrow}\right)&=&-\frac{1-p_0^{\uparrow}}{2-(p_0^{\uparrow}+p_1^{\uparrow})} \log_2(\frac{1-p_0^{\uparrow}}{2-(p_0^{\uparrow}+p_1^{\uparrow})}) \nonumber \\
&&-\frac{1-p_1^{\uparrow}}{2-(p_0^{\uparrow}+p_1^{\uparrow})} \log_2(\frac{1-p_1^{\uparrow}}{2-(p_0^{\uparrow}+p_1^{\uparrow})}) \nonumber
\end{eqnarray}
\normalsize
where $p_0^{\uparrow}=\frac{1}{2}+\frac{M_0}{N}$ ($p_1^{\uparrow}=\frac{1}{2}+\frac{M_1}{N}$) is the probability of finding each spin in state $\ket{\uparrow}$ when the MSS is in state $\ket{S_0}$ ($\ket{S_1}$). Equations \ref{eq:EvDicke} for the special case of Dicke states are reproduced when replacing $p_0^{\uparrow}$ and $p_1^{\uparrow}$ with $\frac{k_0}{N}$ and $\frac{k_1}{N}$, respectively. Plots in figure \ref{fig:Dicke_ent} with the change of variables $\frac{\Delta k}{N}=\Delta p$ and $\frac{\Bar{k}}{N}=\Bar{p}$, represent entropy of entanglement of 
a QA-symmetric state.  
The second order dependency of entanglement reduction to distinctness is observed as,
\small
\begin{eqnarray}
 \Bar{E}_V &=& P_{\uparrow}E_V\left(\phi^{S\uparrow}\right)+P_{\downarrow}E_V\left(\phi^{S\downarrow}\right) \\
 &=& 1-\frac{1}{2\ln{2}}
    \left(\frac{\Delta M}{N}\right)^2\left(1+ 4\left(\dfrac{\Bar{M}}{N}\right)^2 + \frac{1}{6} \left(\frac{\Delta M}{N}\right)^2+... \right) \nonumber
\end{eqnarray}
\normalsize
with $\Delta M = M_0-M_1$ and $\Bar{M}=\frac{M_0+M_1}{2}$. Thus, all the results of the previous section are generalized to the QA-symmetric micro-macro entangled states. This observation proves that among the two characteristics of the magnetization spectra that influence the macroscopicity of a micro-macro entangled state, namely expectation values and SDs of the spectra of $\ket{\psi_0}_{\text{MSS}}$ and $\ket{\psi_1}_{\text{MSS}}$, only the expectation values affect the robustness of entanglement to single particle measurement.

Now lets consider a 
QA-symmetric micro-macro entangled state subject to single particle loss. The state of the qubit and the MSS after losing the $i$th spin in the MSS is,
\begin{equation}
   \rho^{N-1}_{q-\text{MSS}}=\text{Tr}_{i}\left(\ketbra{\phi^S}{\phi^S}_{q-\text{MSS}}\right)
\end{equation}
The entanglement of this mixed state can be quantified with negativity \cite{Vidal02}, defined as the sum of negative eigenvalues of the partial transpose of the state of bipartite system with respect to one party. Negativity ranges from zero \footnote{Note that unless the systems is $2 \times 2$ or $2 \times 3$, in general, zero negativity does not conclude separability \cite{Peres96}, however the reverse is correct, nonzero negativity concludes entanglement.} to $\frac{1}{2}$ for maximally entangled states. Negativity of $\rho^{N-1}_{q-\text{MSS}}$ is a function of  $p_0^{\uparrow}=\frac{1}{2}+\frac{M_0}{N}$ and $p_1^{\uparrow}=\frac{1}{2}+\frac{M_1}{N}$
\small
\begin{widetext}
\begin{eqnarray}
\label{eq:NegS}
    \text{Neg}\left(\rho^{N-1}_{q-\text{MSS}}\right)&=&\frac{1}{2}\left(\sqrt{p_0^{\uparrow}p_1^{\uparrow}}+\sqrt{(1-p_0^{\uparrow})(1-p_1^{\uparrow})}\right) \\
    &=&\frac{1}{2}\left(\sqrt{(\dfrac{1}{2}+\dfrac{M_0}{N})\left(\dfrac{1}{2}+\dfrac{M_1}{N}\right)}+\sqrt{\left(\dfrac{1}{2}-\dfrac{M_0}{N}\right)\left(\dfrac{1}{2}-\dfrac{M_1}{N}\right)}\right) \nonumber
\end{eqnarray}
\end{widetext}
\normalsize
In deriving this relation, we assumed orthogonality of the states in each set of $\{\ket{S_{0\uparrow}^{N-1}},\ket{S_{1\uparrow}^{N-1}}\}$ and $\{\ket{S_{0\downarrow}^{N-1}},\ket{S_{1\downarrow}^{N-1}}\}$, in equations \ref{eq:updatedSym}, 
which follows from separation of the spectra in the original micro-macro entangled state.
Equation \ref{eq:NegS} shows that the robustness of a 
QA-symmetric micro-macro entangled state to single particle loss depends only on the mean value of the magnetization measurement for each state of the superposition and the total number of spins, the SDs of the spectra do not affect the robustness.

\begin{figure}[t!h]
\centering
         \includegraphics[scale=0.5]{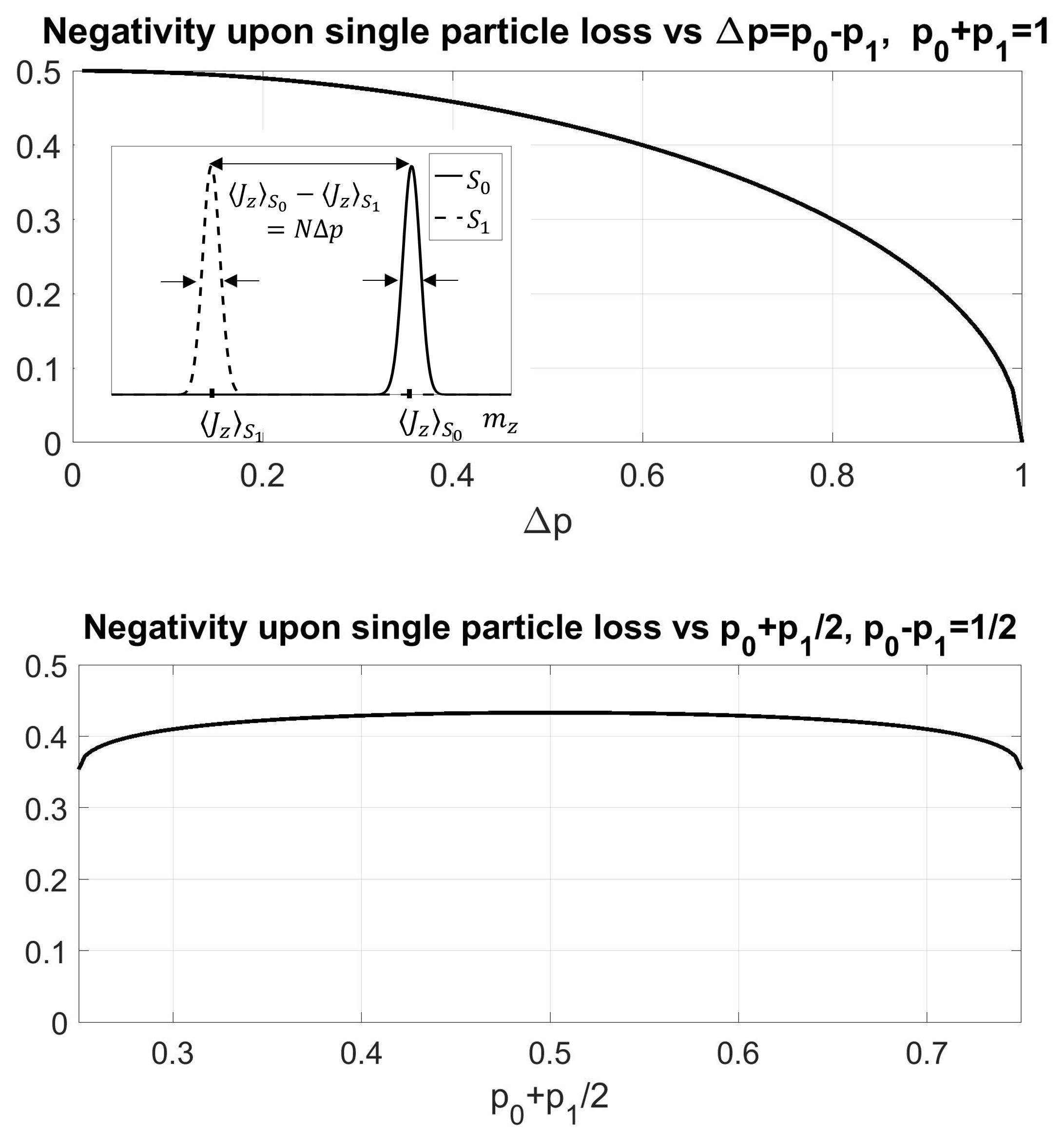}
         \caption{\label{fig:neg} This figure shows entanglement of a QA-symmetric micro-macro entangled state after single particle loss. Entanglement is quantified with the negativity and is plotted as a function of (a) the distinctness normalized to the size of the MSS and (b) the mean of the two spectra normalized to the size of the MSS. The distinctness plays the essential role in the robustness of entanglement to single particle loss while the mean of spectra has a little effect.
         }    
\end{figure}

Figure \ref{fig:neg} plots the negativity of $\rho^{N-1}_{q-\text{MSS}}$ as a function of $\frac{\Delta M}{N}$ and $\frac{\Bar{M}}{N}$ when the other parameter is fixed. Similar to the entropy of entanglement after single particle measurement, the negativity after single particle loss mainly depends on the distinctness of the micro-macro entangled state and the centre of the spectra has little effect.
The plot also shows slow initial drop of the negativity as a function of $\frac{\Delta M}{N}$. Expanding the equation \ref{eq:NegS} in terms of $\frac{\Delta M}{N}$ and $\frac{\Bar{M}}{N}$ shows that this slow drop is  second order in $\frac{\Delta M}{N}$ and the first non-zero term of $\frac{\Bar{M}}{N}$ is $\left(\frac{\Delta M}{N}\right)^2\left(\frac{\Bar{M}}{N}\right)^2$, 
\small
\begin{eqnarray}
     \text{Neg}\left(\rho^{N-1}_{q-\text{MSS}}\right)&=&\frac{1}{2} \\
     &-&\left(\frac{\Delta M}{N}\right)^2\left(\frac{1}{4}+\left(\frac{\Bar{M}}{N}\right)^2+\frac{1}{16}\left(\frac{\Delta M}{N}\right)^2+ ...\right) \nonumber
\end{eqnarray}
\normalsize

In summary, the robustness of QA-symmetric states, which include experimentally relevant states, to single particle noise depends solely on the mean of the spectra of their corresponding mesoscopic states. Moreover, the initial decrease of their entanglement after single particle noise is second order in distinctness. The next section proves that QA-symmetric states are the most robust micro-macro entangled states for a fixed value of distinctness.


\section{Upper Bound of Robustness}
\label{sec:Bounds}
In the this section, we show that for fixed values of the mean of the spectra $M_0$ and $M_1$, a 
QA-symmetric state is the most robust state to single particle measurement and single particle loss and equations \ref{eq:EvS} and \ref{eq:NegS} give the upper bounds on the entanglement subject to single particle noise.


The robustness of a general micro-macro entangled state, presented in equation \ref{eq:mic-mac}, to single particle measurement can be quantified as the average of entanglement after measuring each particle,
\begin{eqnarray}
     \Bar{E}_{V j}&=&p_{j\uparrow}E_V\left(\ket{\phi_{q,\text{MSS}}^{\uparrow j}}\right)+p_{j\downarrow}E_V\left(\ket{\phi_{q,\text{MSS}}^{\downarrow j}}\right)\nonumber \\
     \Bar{E}_V&=&\frac{1}{N}\sum_{j=1}^{N}  \Bar{E}_{V j}
\end{eqnarray}
Here, the states $\ket{\phi_{q,\text{MSS}}^{\uparrow j}}$ and $\ket{\phi_{q,\text{MSS}}^{\downarrow j}}$ are the updated states of the qubit and the MSS after measurement of the jth particle with the outcomes $\ket{\uparrow}$ and $\ket{\downarrow}$, respectively and $p_{j\uparrow}$ and $p_{j\downarrow}$ are the corresponding probabilities.

We answer the following questions: For fixed mean values of magnetization spectra of states  $\ket{\psi_0}_{\text{MSS}}$ and $\ket{\psi_1}_{\text{MSS}}$ what state $\ket{\phi}_{q-\text{MSS}}=\frac{1}{\sqrt{2}}(\ket{0}\ket{\psi_0}_{\text{MSS}}+\ket{1}\ket{\psi_1}_{\text{MSS}})$ is the most robust state to single particle measurement? How robust is that state to single particle measurement?
In particular, we find the upper bound of the average entropy of entanglement when fixing the mean values of magnetization spectra of states  $\ket{\psi_0}_{\text{MSS}}$ and $\ket{\psi_1}_{\text{MSS}}$ to be $M_0$ and $M_1$. 
The average entropy of entanglement upon measuring jth spin in the MSS, can be written as, 
\begin{widetext}
\begin{eqnarray}
\label{eq:EVj}
     \Bar{E}_{Vj}&=&-\frac{1}{2}\left(p_{0j}^{\uparrow} \log_2\left(\frac{p_{0j}^{\uparrow}}{p_{0j}^{\uparrow}+p_{1j}^{\uparrow}}\right)+p_{1j}^{\uparrow}\log_2\left(\frac{p_{1j}^{\uparrow}}{p_{0j}^{\uparrow}+p_{1j}^{\uparrow}}\right)\right) \\
     && -\frac{1}{2}\left((1-p_{0j}^{\uparrow}) \log_2\left(\frac{1-p_{0j}^{\uparrow}}{2-p_{0j}^{\uparrow}-p_{1j}^{\uparrow}}\right)+(1-p_{1j}^{\uparrow})\log_2\left(\frac{1-p_{1j}^{\uparrow}}{2-p_{0j}^{\uparrow}-p_{1j}^{\uparrow}}\right)\right) \nonumber
\end{eqnarray}
\end{widetext}
 
 where $p_{0j}^{\uparrow}$ and $p_{1j}^{\uparrow}$ are the probabilities of  finding spin $j$ in state $\ket{\uparrow}$ when the MSS is in state $\ket{\psi_0}_{\text{MSS}}$ and $\ket{\psi_1}_{\text{MSS}}$, respectively.
Total magnetization of $M_0$ and $M_1$, requires, 
\begin{eqnarray}
\label{eq:conds}
     \sum_{j=1}^N p_{0j}^{\uparrow}&=&M_0+\frac{N}{2} \\
     \sum_{j=1}^N p_{1j}^{\uparrow}&=&M_1+\frac{N}{2} \nonumber
\end{eqnarray}
We prove that the maximum of $ \Bar{E}_V=\frac{1}{N}\sum_{j=1}^{N}  \Bar{E}_{V j}$ happens when 
$p_{0j}^{\uparrow} \equiv p_{0}^{\uparrow}=\frac{M_0}{N}+\frac{1}{2}$ and $p_{1j}^{\uparrow} \equiv p_{1}^{\uparrow}=\frac{M_1}{N}+\frac{1}{2}$, for all $j=1,2, ..., N$,
which corresponds to a 
QA-symmetric micro-macro entangled state.
Thus, the maximum of entropy of entanglement is given by setting $p_{0j}^{\uparrow}$ and $p_{1j}^{\uparrow}$ to $p_{0}^{\uparrow}=\frac{M_0}{N}+\frac{1}{2}$ and $p_{1}^{\uparrow}=\frac{M_1}{N}+\frac{1}{2}$ in equation \ref{eq:EVj}.
\begin{widetext}
\begin{eqnarray}
\label{eq:EVMax}
     \Bar{E}_{V}^{\text{max}}&=&-\frac{1}{2}\left((\frac{1}{2}+\frac{M_0}{N}) \log_2\left(\frac{\frac{1}{2}+\frac{M_0}{N}}{1+\frac{M_0}{N}+\frac{M_1}{N}}\right)+(\frac{1}{2}+\frac{M_1}{N})\log_2\left(\frac{\frac{1}{2}+\frac{M_1}{N}}{1+\frac{M_0}{N}+\frac{M_1}{N}}\right)\right) \\
     && -\frac{1}{2}\left((\frac{1}{2}-\frac{M_0}{N}) \log_2\left(\frac{\frac{1}{2}-\frac{M_0}{N}}{1-\frac{M_0}{N}-\frac{M_1}{N}}\right)+(\frac{1}{2}-\frac{M_1}{N})\log_2\left(\frac{\frac{1}{2}-\frac{M_1}{N}}{1-\frac{M_0}{N}-\frac{M_1}{N}}\right)\right) \nonumber
\end{eqnarray}
\end{widetext}
The detailed proof is given in Appendix \ref{ap:1}.
Here I give an overview of the proof. 
We search for the maximum of $\Bar{E}_V$ in the region where $p_{0j}^{\uparrow}\in [0,1]$, $p_{1j}^{\uparrow}\in [0,1]$ for all $j=1, 2, ..., N$, and they follow the conditions in equations \ref{eq:conds}. The function $\Bar{E}_V$ has 2N variables among which 2N-2 are independent. 
To find its maxima, we first find its derivatives with respect to the $2N-2$ variables and set those to zero. The zero derivatives happen when $p_{01}^{\uparrow}=p_{02}^{\uparrow}=...=p_{0N}^{\uparrow}$ and $p_{11}^{\uparrow}=p_{12}^{\uparrow}=...=p_{1N}^{\uparrow}$. 

Next, 
 we find the Hessian matrix at this point and show that all its eigenvalues are negative which proves that this point is a local maximum. Since the function is continuous and this point is the only point with zero-derivative, it is also a global maximum of  $\Bar{E}_V$. Thus, the maximum of  $\Bar{E}_V$ happens for collective 
QA-symmetric states and its value is given with the equation \ref{eq:EVMax}. Similarly, it can be shown that maximum average negativity, $\Bar{Neg}^{\text{max}}$, of a micro-macro entangled state 
upon single particle loss happens for 
QA-symmetric states and its value is given by equation \ref{eq:NegS}.
%
%

To summarize, the robustness of micro-macro entangled states to single particle noise solely depends on the mean of the spectra of the corresponding states of the MSS and their symmetry with respect to measurement of the constituent two-level systems along the quantization axis. Other factors such as the standard deviation and higher moments of the magnetization spectra and entanglement or separability of the MSS's states, $\ket{\psi_0}_{\text{MSS}}$ and $\ket{\psi_1}_{\text{MSS}}$, 
do not affect the robustness.
With fixed mean of the magnetization spectra, 
QA-symmetric states, defined as the states satisfying $\left< \sigma^i_z\right>_{\psi_0}=\left< \sigma^j_z\right>_{\psi_0}$ and $\left< \sigma^i_z\right>_{\psi_1}=\left< \sigma^j_z\right>_{\psi_1}$ for all pairs of spins, $i$ and $j$, are the most robust states. The decrease of the entanglement entropy and negativity for such a state is of the second order of $\frac{\Delta M}{N}$ and next terms are of the order of $\left(\frac{\Delta M}{N}\right)^2 \left(\frac{\Bar{M}}{N}\right)^2$ and $\left(\frac{\Delta M}{N}\right)^4$.




\section{Summary and Conclusion}
\label{sec:Conclusion}
We quantified the robustness of micro-macro entangled states 
involving mesoscopic spin systems to single particle events including single particle measurement and single particle loss. 
Specifically, we looked at how the reduction of entanglement, quantified with entanglement entropy and negativity, scales with the macroscopic distinctness 
of the micro-macro entangled state and what other properties of the state play role in its robustness.

The results can be summarized as,
\begin{itemize}
    \item As the macroscopic distinctness increases relative to the size of the mesoscopic system, the micro-macro entangled state becomes more fragile to single particle events, as expected. 
    \item Micro-macro entangled states, whose macroscopic party poses symmetric states with respect to the 
    measurement axis, such as the experimentally relevant states \cite{Mirkamali20}, symmetric separable states, Dicke states or superposition of Dicke states, are proved to be the most robust states to microscopic events. Their entanglement after single particle noise gives an upper bound on the robustness. Thus, 
    the most robust micro-macro entangled states 
    have information distributed uniformly among all particles in the mesoscopic system.
     \item 
     The initial drop of the entanglement's upper bound is slow and second order in the distinctness. 
    This result implies that there is a regime where 
    bipartite entangled states between a a qubit and a mesoscopic system are both macroscopic and robust to single particle noise.
    To have a robust micro-macro entangled state, 
    the distinctness needs to be small compared to the size of the mesoscopic system and large compared to the standard deviations of the spectra of the corresponding MSS's states.
    \item The robustness of symmetric micro-macro entangled states depend on the the mean of the spectra of each of the two states of the mesoscopic system, while the standard deviation and higher moments of the spectra do not matter.
\end{itemize}

These results advance our understanding of the noise effects on quantum behaviour at mesoscopic scales and quantum to classical transition in many-body systems. Furthermore, they help identifying experimental requirements for realization of micro-macro entangled spin states as an emergent resource for quantum information processing.
In this work, we analysed single particle noise. The effect of noise on all particles is subject to future studies.


\section{Acknowledgements}
This work was supported by the Canada First Research Excellence Fund (CFREF), the Canadian Excellence Research Chairs (CERC 215284) program, the Natural Sciences and Engineering Research Council of Canada (NSERC RGPIN-418579) Discovery program, and the Province of Ontario.

\appendix


\section{Upper bound on Entanglement upon single particle measurement}
\label{ap:1}
Here, we prove that among all micro-macro entangled states with fixed collective magnetization of the corresponding MSS's states, $\left<J_z\right>_{\psi_0}=M_0$ and $\left<J_z\right>_{\psi_1}=M_1$, 
collective 
QA-symmetric micro-macro entangled states are the most robust ones to single particle noise and the entanglement entropy of these states after single particle loss is given by equation \ref{eq:EVMax}.

The average entropy of entanglement of a micro-macro entangled state upon losing one particle in the MSS is,

\begin{widetext}
\begin{eqnarray}
\label{eq:EVAvg}
     \Bar{E}_{V}=\frac{1}{N}\sum_{j=1}^{N}&-&\frac{1}{2}\left(p_{0j}^{\uparrow} \log_2\left(\frac{p_{0j}^{\uparrow}}{p_{0j}^{\uparrow}+p_{1j}^{\uparrow}}\right)+p_{1j}^{\uparrow}\log_2\left(\frac{p_{1j}^{\uparrow}}{p_{0j}^{\uparrow}+p_{1j}^{\uparrow}}\right)\right) \\
     && -\frac{1}{2}\left((1-p_{0j}^{\uparrow}) \log_2\left(\frac{1-p_{0j}^{\uparrow}}{2-p_{0j}^{\uparrow}-p_{1j}^{\uparrow}}\right)+(1-p_{1j}^{\uparrow})\log_2\left(\frac{1-p_{1j}^{\uparrow}}{2-p_{0j}^{\uparrow}-p_{1j}^{\uparrow}}\right)\right) \nonumber
\end{eqnarray}
\end{widetext}
where $p_{0j}^{\uparrow}$ and $p_{1j}^{\uparrow}$ are the probabilities of finding spin $j$ in the MSS in state $\uparrow$ when the MSS is in state $\ket{\psi_0}_{\text{MSS}}$ and $\ket{\psi_1}_{\text{MSS}}$, respectively. To find the extremum, one can find the derivative of $\Bar{E}_{V}$ with respect to each of its variables $p_{0j}^{\uparrow}$ and $p_{1j}^{\uparrow}$, $j=1,2, ...,N-1$ and set those to zero. The probabilities  $p_{0N}^{\uparrow}=M_0+\frac{N}{2}-\sum_{j=1}^{N-1} p_{0j}^{\uparrow}$, $p_{1N}^{\uparrow}=M_1+\frac{N}{2}-\sum_{j=1}^{N-1} p_{1j}^{\uparrow}$ are not independent variables, thus using chain rule, $\frac{d \Bar{E_V}}{d p_{0j}^{\uparrow}}=\frac{\partial \Bar{E_V}}{\partial p_{0j}^{\uparrow}}+\frac{\partial \Bar{E_V}}{\partial p_{0N}^{\uparrow}}\frac{\partial p_{0N}^{\uparrow}}{\partial p_{0j}^{\uparrow}}$. The derivatives $\frac{d \Bar{E_V}}{d p_{0j}^{\uparrow}}$ and $\frac{d \Bar{E_V}}{d p_{1j}^{\uparrow}}$ after simplifying are,
\begin{widetext}
\begin{eqnarray}
\label{eq:EVAvg}
     \frac{d \Bar{E_V}}{dp_{0j}^{\uparrow}}&=&-\frac{1}{2}\left( \log_2\frac{p_{0j}^{\uparrow}}{p_{0j}^{\uparrow}+p_{1j}^{\uparrow}}-\log_2\frac{1-p_{0j}^{\uparrow}}{2-p_{0j}^{\uparrow}-p_{1j}^{\uparrow}}-\log_2\frac{p_{0N}^{\uparrow}}{p_{0N}^{\uparrow}+p_{1N}^{\uparrow}}-\log_2\frac{1-p_{0N}^{\uparrow}}{2-p_{0N}^{\uparrow}-p_{1N}^{\uparrow}}\right) \\
     \frac{d \Bar{E_V}}{d p_{1j}^{\uparrow}}&=& -\frac{1}{2}\left( \log_2\frac{p_{1j}^{\uparrow}}{p_{0j}^{\uparrow}+p_{1j}^{\uparrow}}-\log_2\frac{1-p_{1j}^{\uparrow}}{2-p_{0j}^{\uparrow}-p_{1j}^{\uparrow}}-\log_2\frac{p_{1N}^{\uparrow}}{p_{0N}^{\uparrow}+p_{1N}^{\uparrow}}-\log_2\frac{1-p_{1N}^{\uparrow}}{2-p_{0N}^{\uparrow}-p_{1N}^{\uparrow}}\right) \nonumber
\end{eqnarray}
\end{widetext}
Setting both equations to zero gives, $p_{0j}^{\uparrow}=p_{0N}^{\uparrow}$ and $p_{1j}^{\uparrow}=p_{1N}^{\uparrow}$ for any $j=1,2,...,N-1$. Thus the zero-derivative of $ \Bar{E_V}$ happens at $p_{01}^{\uparrow}=p_{02}^{\uparrow}=...=p_{0N}^{\uparrow}$ and $p_{11}^{\uparrow}=p_{12}^{\uparrow}=...=p_{1N}^{\uparrow}$. Next, we show that this zero-derivative point is a local maximum by finding the second derivative matrix, called the Hessian matrix, and showing all its eigenvalues are negative at this point. Noting that this point is the only point with zero derivative proves that it is a global maximum.
With the ordering of parameters, $\{p_{01}^{\uparrow},p_{02}^{\uparrow},...p_{0N-1}^{\uparrow},p_{11}^{\uparrow},p_{12}^{\uparrow}, ..., p_{1N-1}^{\uparrow}\}$, the Hessian matrix at the zero-derivative point has the form,
\begin{equation}
    H_{\Bar{E_V}}=-\frac{1}{2\ln{2}}\begin{pmatrix}
    2a&a& ... & a &2c&c&...&c\\
    a&2a& ...& a & c& 2c& ...&c\\
    .&.& ...& .& .&.& ...& .  \\
     .&.& ...& .& .&.& ...& .  \\
      .&.& ...& .& .&.& ...& .  \\
  a&a& ...& 2a & c& c& ...&2c\\
      2c&c& ... & c &2b&b&...&b\\
    c&2c& ...& c & b& 2b& ...&b\\
    .&.& ...& .& .&.& ...& .  \\
     .&.& ...& .& .&.& ...& .  \\
      .&.& ...& .& .&.& ...& .  \\
     c&c& ...& 2c & b& b& ...&2b
    \end{pmatrix}
\end{equation}
with 
\begin{eqnarray}
a&=&\frac{p_1}{p_0(p_0+p_1)}+\frac{1-p_1}{(1-p_0)(2-p_0-p_1)} \\
b&=&\frac{p_0}{p_1(p_0+p_1)}+\frac{1-p_0}{(1-p_1)(2-p_0-p_1)} \nonumber \\
c&=&-\frac{1}{p_0+p_1}-\frac{1}{2-p_0-p_1} \nonumber
\end{eqnarray}
where $p_0=\frac{M_0}{N}+\frac{1}{N}$ and $p_1=\frac{M_1}{N}+\frac{1}{N}$.
The eigenvalues of this matrix are $-\frac{1}{4\ln{2}}\left(a+b-\sqrt{(a-b)^2+4c^2}\right)$ and $-\frac{1}{4\ln{2}}\left(a+b+\sqrt{(a-b)^2+4c^2}\right)$ repeated $N-2$ times and  $-\frac{N}{4\ln{2}}\left(a+b-\sqrt{(a-b)^2+4c^2}\right)$ and $-\frac{N}{4\ln{2}}\left(a+b+\sqrt{(a-b)^2+4c^2}\right)$. The value of $(a+b)^2-\left((a-b)^2+4c^2\right)=4\frac{(p_0-p_1)^2(1-p_0-p_1)^2}{(1-p_0)p_0(1-p_1)p_1(2-p_0-p_1)(p_0+p_1)}$ is always positive given that $0<p_0<1 (M_0\neq \pm\frac{N}{2})$, $0<p_1<1  (M_1\neq \pm\frac{N}{2})$, $p_0\neq p_1 (M_0\neq M_1)$ (not boundary points) \footnote{This condition is always satisfied for a micro-macro entangled state},  and $p_0+p_1\neq 1 (M_0\neq -M_1)$ \footnote{For this special case, some eigenvalues of the Hessian matrix are zero, more information is needed to confirm the maximal point.}. 
When these conditions are satisfied, all eigenvalues are negative and the equal probabilities correspond to local and global maximum.

\bibliographystyle{apsrev4-1} 

\bibliography{references}




\end{document}